\documentclass{PoS}

\usepackage{amssymb}
\usepackage{amsmath}
\usepackage{graphicx}  
\usepackage{epsfig}
\newcommand{\eps}{\epsilon}
\def\dsp{\displaystyle}
\title{The structure of the nucleon from generalized parton distributions}

\ShortTitle{Generalized Parton Distributions}

\author{\speaker{B. Pasquini}, S. Boffi
\\
Dipartimento di Fisica Nucleare e Teorica, Universit\`{a} degli Studi di Pavia, and
Istituto Nazionale di Fisica Nucleare, Sezione di Pavia, I-27100 Pavia, Italy
       \\
        E-mail: \email{pasquini@pv.infn.it}

}
\abstract{Generalized parton distributions (GPDs) have become a standard  QCD tool for
 analyzing and parametrizing the non perturbative parton structure of hadron
 targets. 
GPDs  might be viewed as non-diagonal overlaps of light-cone wave functions and 
offer the opportunity to study the partonic content of the nucleon 
 from a new perspective,
 allowing one to study the interplay between longitudinal and 
transverse partonic degrees of freedom.
In particular, we will review some of the new information 
encoded in the GPDs through the definition of impact-parameter dependent
parton distributions and form factors of the energy-momentum tensor, by 
exploiting different dynamical models for the nucleon state.
}

\FullConference{LIGHT CONE 2008 Relativistic Nuclear and Particle Physics\\
		 July 7-11  2008\\
		 Mulhouse, France}

\begin{document}

\newcommand{\tvec}[1]{\mbox{\boldmath{$#1$}}}
\newcommand{\svec}[1]{\mbox{\boldmath{$\scriptstyle #1$}}}
\newcommand{\lrpartial}{\raisebox{0.1em}{$
\stackrel{\raisebox{-0.03em}{$\scriptstyle\leftrightarrow$}}{\partial}$}{}}

\renewcommand{\slash}[1]{#1 \hspace{-0.45em} / }

\def\ket#1{\hbox{$\vert #1\rangle$}}   
\def\bra#1{\hbox{$\langle #1\vert$}}   
\def\be{\begin{equation}}
\def\ee{\end{equation}}
\def\bea{\begin{eqnarray}}
\def\eea{\end{eqnarray}}

\def\oneh{{\textstyle {1\over 2}}}
\def\onet{{\textstyle {1\over 3}}}
\def\smoneh{{\scriptstyle {1\over 2}}}
\def\onesix{{\textstyle {1\over 6}}}
\def\oneq{{\textstyle {1\over 4}}}
\def\treh{{\textstyle {3\over 2}}}
\def\treq{{\textstyle {3\over 4}}}
\def\oneight{{\textstyle {1\over 8}}}
\def\onesq{{\textstyle {1\over \sqrt{2}}}}

\def\Re{\hbox{\rm Re\,}}
\def\Im{\hbox{\rm Im\,}}

\section{Introduction}

Generalized Parton Distributions (GPDs) have become a standard QCD tool for
 analyzing and parametrizing the non perturbative parton structure of hadron
 targets, for reviews see~\cite{GPVdH,Diehlrep,Jiarnps,BR05,BP:review}.
GPDs have been introduced in the past in different contexts 
(see, e.g., \cite{Dittes88,Muller94}), but have raised a large interest 
in the hadron community only when their importance was stressed in studies of deeply virtual Compton scattering (DVCS)~\cite{Ji97, Radyushkin96a,Ji97a} 
and hard meson production~\cite{Radyushkin96b} in connection with the 
possibility of factorizing their contribution and gaining 
information on the spin structure of the nucleon~\cite{Ji97}. 
Being defined in terms of nondiagonal matrix elements of the same correlation functions entering the definition of the parton distribution (PDs), 
GPDs reveal the partonic content of hadrons from a complementary perspective.
They
do not represent any longer a probability,
 but rather the interference between amplitudes describing 
different parton configurations of the nucleon so that they give access to momentum correlations of partons in the nucleon.
Furthermore, the finite momentum transfer to the proton makes a second space-time structure
 of the process possible, and after the Fourier transform in the impact-parameter space allows one to define spin-dependent densities which describe
 how partons are spatially distributed in the 
transverse plane~\cite{Burkardt00a}.

The GPDs can also be viewed as the generating functions for the form factors 
of the twist-two operators  governing the interaction mechanisms of hard 
processes in the deep inelastic regime.
 These generalized form factors do not couple directly to any known 
fundamental interactions, but can be studied indirectly looking at moments 
of the GPDs. The most peculiar example are the form factors of the energy-momentum tensor, which give information about the spatial distribution
of energy, angular momentum and forces experienced by 
quarks and gluons 
inside hadrons.

After summarizing in sect. 2  definition and basic  properties
of  GPDs,  
in sect. 3 we
show results for the GPDs in the impact-parameter space using a light-cone quark model and discussing in particular the correlations of spin and 
orbital angular momentum of the quarks in the nucleon.
Finally, in sect. 4 we review some results  for
the form factors of the energy-momentum 
tensor in the framework of the chiral-quark soliton model. 
\section{Definition and basic properties of GPDs}

Parton distributions are defined in terms of matrix elements of 
light-cone bilocal operators between proton states of equal momenta. 
In general,  with initial (final) momentum $p$ ($p'$)  
and helicity $\lambda_N$ ($\lambda_N'$) one defines a set of quark generalized 
quark distributions for a hadron with spin $\frac{1}{2}$ 
\begin{eqnarray}
  \langle p',\lambda_N'|\, 
{\cal O}^{\Gamma}(x,{\bf 0}_\perp)
  \,|p,\lambda_N \rangle ,
\label{eq:def}
\end{eqnarray}
with
\begin{equation}
{\cal O}^\Gamma(x,{\bf 0}_\perp)=
 \int \frac{d z^-}{4\pi}\, e^{ix P^+ z^-}
     \bar{\psi}_q(-\frac{z^-}{2},{\bf 0}_\perp)\, 
     \Gamma \psi_q(\frac{z^-}{2},{\bf 0}_\perp).\, 
\label{eq:correl}
\end{equation}
In Eq.~(\ref{eq:correl}) $P=(p+p')/2$,
and the operator $\Gamma$  is a matrix in Dirac space which
selects different spin polarizations of the quark fields.
For three particular  matrices $\Gamma$ one can classify 
eight leading twist GPDs: 
{\em i}) two unpolarized quark GPDs,
$H^q$ and $E^q$, for $\Gamma=\gamma^+$;
{\em ii}) two longitudinally 
polarized quark GPDs, $\widetilde H^q$ and $\widetilde E^q$,
for $\Gamma=\gamma^+\gamma_5$;
{\em iii})  four quark 
chiral-odd GPDs for $\Gamma=i\sigma^{i+}\gamma_5$, i.e.
$H_T^q$ and $E_T^q,$ which involve the density 
operator for transversely polarized quarks,
and
$\widetilde H_T^q$ and $\widetilde E_T^q$,  defined in terms of 
a quark operator which flips 
the transverse
spin of the quark.
The tilded distributions correspond to matrix elements between nucleon states
with flip of the polarization, while the untilded distributions refer to
no-flip of the nucleon polarization. 
Analogous definitions hold for the gluon GPDs.

Because of Lorentz invariance the eight GPDs
 can only depend on three kinematical variables, 
i.e. the (average) quark longitudinal momentum fraction $x=k^+/P^+$, the invariant momentum square $t=\Delta^2\equiv (p'-p)^2,$ 
and the skewness parameter $\xi$ given by
$
\xi = -\Delta^+/(2P^+).$
In addition, there is an implicit scale dependence in the definition of GPDs corresponding to the factorization scale $\mu^2$ used to separate the (universal) matrix element defining a GPD inside the entire amplitude describing the process under study.

In the forward case, $p=p'$, both $\Delta$ and $\xi$ are zero. 
In this case the  functions $H^q$, $\widetilde H^q$ and $H_T^q$ reduce to the usual DIS parton distribution functions, i.e. 
the quark density, helicity and transversity distributions, respectively.
No corresponding relations exist for the functions $E^q$, $\widetilde E^q$, $E_T^q$ and $\widetilde H_T^q$, because in the forward limit they decouple in their defining equations. However, they do not vanish. In particular, $E^q(x,0,0)$ carries important information about the quark orbital angular momentum.
In contrast, $\widetilde E_T^q(x,0,0)$ vanishes identically being an odd function of $\xi$ by time reversal symmetry.

Moments in the momentum fraction $x$ play an important role in the theory 
of GPDs. Weighting Eq.~(\ref{eq:def}) with integer powers of $x$ and integrating over $x$, the correlation function ${\cal O}^\Gamma$ reduces to a local operator and the corresponding matrix elements can be parametrized in terms of generalized form factors (GFFs) related to Mellin moments of the GPDs.
More specifically, one has
\begin{eqnarray}
\int_{-1}^{+1} dx\,x^{n-1}
\left[
\begin{matrix}
H^q(x,\xi,t) \\
E^q(x,\xi,t) \\
\end{matrix}
\right]
=\sum_{i=0 \atop \scriptstyle{\rm even}}^{n-1}
\left[
\begin{matrix}
A^q_{n,i}(t) \\
B_{n,i}^q(t) \\
\end{matrix}
\right]
(2 \xi)^{i}
\pm {\rm Mod}(n+1,2) C_{n}^q(t)\;(2 \xi)^n
\label{eq:mellin1}
\end{eqnarray}
 for the unpolarized GPDs;


\begin{eqnarray}
\int_{-1}^{+1} dx\,x^{n-1}
\widetilde H^q(x,\xi,t) 
=\sum_{i=0 \atop \scriptstyle{\rm even}}^{n-1}(2\xi)^i
\widetilde  A^q_{n,i}(t),\quad
\int_{-1}^{+1} dx\,x^{n-1}
\widetilde E^q(x,\xi,t) 
=\sum_{i=0 \atop \scriptstyle{\rm even}}^{n-1}(2\xi)^i
\widetilde   B_{n,i}^q(t) 
\label{eq:mellin2}
\end{eqnarray}
for the polarized GPDs;


\begin{eqnarray}
\int_{-1}^{+1} dx\,x^{n-1}
H_T^q(x,\xi,t) 
=\sum_{i=0 \atop \scriptstyle{\rm even}}^{n-1}
(2\xi)^iA^q_{Tn,i}(t),\quad
\int_{-1}^{+1} dx\,x^{n-1}
E_T^q(x,\xi,t) 
=\sum_{i=0 \atop \scriptstyle{\rm even}}^{n-1}
(2\xi)^i B_{Tn,i}^q(t),
\end{eqnarray}
\begin{eqnarray}
\label{eq:chiral-tilde}
\int_{-1}^1 dx\, x^{n-1} \widetilde{H}^q_T(x,\xi,t) =
  \sum_{i=0 \atop \scriptstyle{\rm even}}^{n-1} (2\xi)^i 
        \widetilde{A}^q_{Tn,i}(t) ,
\quad
\int_{-1}^1 dx\, x^{n-1} \widetilde{E}^q_T(x,\xi,t) =-
  \sum_{i=0 \atop \scriptstyle{\rm odd}}^{n-1} (2\xi)^i 
        \widetilde{B}^q_{Tn,i}(t)
\label{eq:mellin3}
\end{eqnarray}
for the chiral-odd GPDs.

For the lowest moment $n=1$ in Eqs.~(\ref{eq:mellin1})-(\ref{eq:mellin2})
one finds $A^q_{1,0}(t)=F^q_1(t)$, $B^q_{1,0}(t)=F_2^q(t)$,
$\widetilde A^q_{1,0}(t)=g^q_A(t)$ and $\widetilde B^q_{1,0}(t)=g^q_P(t)$ where 
$F_1^q$, $F_2^q$, $g_A^q$ and $g_P^q$ are the quark contribution to the
Dirac, Pauli, axial and induced pseudoscalar form factors, respectively.
Furthermore, in the chiral-odd sector one finds
$A^q_{T1,0}(t)=g^q_T(t)$ and $2 \widetilde A^1_{T1,0}(0)+B^q_{T1,0}(0)=\kappa_T^q$,
where  $g_T^q$ is the quark tensor form factor and $ \kappa_T^q$ 
describes 
how far and in which direction the average position of quarks 
with spin in the $\hat x$-direction is shifted in the 
$\hat y$-direction in an unpolarized nucleon~\cite{Burkardt05b}. 

The second Mellin moments of unpolarized GPDs can be related to the form factors
of the energy-momentum tensor (EMT) of QCD by~\cite{Polyakov03}
\begin{eqnarray}
A^q_{2,0}(t)=M_2^q(t)+\frac{4}{5}d_1^q(t)\xi^2,\qquad
B^q_{2,0}(t)=2J^q(t)-M_2^q(t)-
\frac{4}{5}d_1^q(t)\xi^2.
\end{eqnarray}
The form factor $M_2(t)$ at $t=0$ reduces to the second Mellin moment of unpolarized parton distributions accessible in inclusive deep inelastic scattering,
and represents the fraction
 of the nucleon momentum carried by quarks.
The form factor $d_1(t)$ provides information on the distribution of
strong forces in the nucleon, similarly as the electromagnetic form factors contain information about the electric charge distribution. 
The form factor $J^q(t)$ is relevant for the spin structure 
of the nucleon thanks to the so called Ji's sum rule\cite{Ji97,Ji97a}
\be
\label{eq:jisumrule}
J^q(t=0)=
\langle J^i_{q}\rangle = S^i  \left[A^{q}_{2,0}(0) + B^{q}_{2,0}(0)\right], \quad
\ee 
where $\langle J^i_{q}\rangle$ is the total angular momentum along the direction $\hat i$ 
carried by quarks and antiquarks in a proton with spin $S^i$.
In the case of a proton polarized in the positive $\hat z$-direction,
one can further split the Ji's sum rules  
into spin and orbital angular momentum parts, i.e.
$\displaystyle{\frac{1}{2}}=\displaystyle{\frac{1}{2}}\Sigma^q+L^q$
where the contribution from the quark spin can be obtained from the
moments of the usual polarized quark densities, i.e.
$\Sigma^q=\widetilde A^q_{1,0}(t=0)$.
The orbital angular momentum has recently been calculated 
in lattice simulations by the LHPC~\cite{Hagler07} and QCDSF~\cite{QCDSF04} collaborations.
These calculations are
for pion masses as low as 350 MeV and volume as large as (3.5 fm)$^3,$ 
providing results in the $\overline{{\mbox MS}}$ renormalization scheme.  Extrapolation
to the physical pion mass requires a combination of full QCD lattice and 
Chiral Perturbation Theory~\cite{Dorati}.
 Two remarkable features are found. 
The first is that the magnitude of the orbital angular momentum contributions 
of the up and down quarks are separately sizable, 
$L^u\approx -L^d\approx 0.30$, yet they cancel nearly completely 
at all pion masses, $L^{u+d}\approx 0$~\cite{Hagler07,brommel}, indicating that
the total angular momentum of quarks in the nucleon is of the same size as the
 quark spin contribution.
The  second is the close cancellation between the orbital and spin 
contributions of the down quarks for all pion masses, 
$J^d\approx 0$~\cite{Hagler07,brommel}. 
However, before drawing definite conclusions, 
one should be aware that these results do not include
the contributions from disconnected graphs.
Such contributions cancel in the difference of $u$ and $d$ quark 
distributions but may well be important in their sum.
\newline
From the experimental side,
first model-dependent constraints on the angular momentum were 
extracted from recent DVCS data.
The analysis was performed comparing  data
from HERMES~\cite{Hermes08} on transverse-target and beam-charge asymmetries, 
and cross section data  from JLab~\cite{JLab07}
with various GPD-models having
$J^q$ as free parameters~\cite{VGG,dual}.
Although the  extracted values are strongly model-dependent, such analysis 
show for the first time that DVCS data have indeed the potential to provide
quantitative information about the spin content of the nucleon.

Whereas Eq.~(\ref{eq:jisumrule}) provides the angular momentum carried by the quarks regardless of their spin,
one can also investigate how much each quark polarization component contributes to $\langle J^i_{q}\rangle.$ 
Taking for example the case of transverse polarization in the $\hat x$ direction,
one can 
decompose  $J^x_{q}$
with respect to quarks of definite transversity,
i.e. $J^x_{q}=J^x_{q,+\hat x}+J^x_{q,-\hat x}$
where
$J^x_{q,\pm\hat x}$ corresponds
to the angular momentum in the $\hat x$ direction
carried by quarks with transverse polarization in the $\pm\hat x$-direction~\cite{Burkardt05b}.
The transversity components $J^x_{q,\pm\hat x}$ can be related to 
the second Mellin moments of GPDs as follows
\begin{equation}
\langle J^x_{q,\pm\hat x}\rangle = 
\frac{S^x}{2}\left[A^{q}_{2,0}(0) + B^{q}_{2,0}(0)\right]
\pm
\frac{1}{4} \left[ A^q_{T2,0}(0) + 2 \widetilde A^q_{T2,0}(0) 
+ B^q_{T2,0}(0)\right].
\label{eq:spin}
\end{equation}
In the case of transversely polarized nucleon, 
$J^x_{q,+\hat x}+J^x_{q,-\hat x}$ gives the sum rule~(\ref{eq:jisumrule}), 
while for unpolarized target
only the second term in Eq.~(\ref{eq:spin})
 contributes.
Thanks to this decomposition of the angular momentum in transversity components
one can gain information on the correlation between the transverse spin and the transverse angular momentum carried by the quarks in an unpolarized target.
Note that the same linear combination
of GPDs ($2\widetilde H_T(x,0,0)+E_T(x,0,0)$) that appears in Eq.~(\ref{eq:spin})
also describes the transverse displacement of quarks with a given transversity in an unpolarized target relative to the center of momentum (see sect. 3).

\section{Spin densities in the impact-parameter space}

A convenient way to make explicit which kind of information on hadron 
structure is contained in the GPDs is the representation in terms 
of overlaps 
of light-cone wave functions (LCWFs) which are the probability amplitudes
to find a given $N$-parton configuration in the Fock-space expansion 
of the hadron state.
In the following, we will confine our analysis to the three-quark sector, 
by truncating the light-cone expansion of the nucleon state to the minimal
Fock-space configuration and adopting
a light-cone constituent quark model (CQM) which has been successfully 
applied in the calculation of the electroweak properties of 
the nucleon~\cite{PBff}.
\newline
As outlined in Ref.~\cite{BPT},
the starting point is the three-quark wave function obtained as solution
of the  Schr\"odinger-like eigenvalue equation in the instant-form dynamics.
The corresponding solution in light-cone dynamics is obtained through the
unitary  transformation represented by product of Melosh rotations acting
on the spin of the individual quarks.
In particular, the instant-form wave function is constructed as a product of a momentum wave function which is spherically symmetric and invariant under permutations, and a spin-isospin wave function which is uniquely determined by SU(6)
symmetry requirements.
By applying the Melosh rotations, the Pauli spinors of the quarks in the
 nucleon rest frame are converted to the light-front spinors.
The relativistic spin effects are evident in the presence
of spin-flip terms in the Melosh rotations which generate non-zero orbital angular momentum components
and non-trivial correlations between spin and transverse momentum of the 
quarks. On the other hand,
the momentum-dependent wave function keeps the original functional form, with 
instant-form coordinates rewritten in terms of light-front coordinates.
Model results for the non-polarized, polarized and chiral-odd GPDs in the 
momentum space have been presented in Ref.~\cite{BPT,PBcloud},
with an extension of the model to include the contribution from 
higher-Fock state components in Ref.~\cite{PBcloud}. 
Here we focus on the GPDs in the impact-parameter space,
which give complementary information to the transverse momentum dependent parton distributions on
the spin-spin and spin-orbit correlations of quarks in the nucleon~\cite{PCB:TMD}.

When $\xi=0$ and $x>0$, by a two-dimensional Fourier transform to 
impact-parameter space, GPDs can be interpreted as densities of quarks with 
longitudinal momentum fraction $x$ and transverse location $\tvec b$ with 
respect to the nucleon center of momentum~\cite{Burkardt00a,Burkardt03}. 
Depending on the polarization of both the active quark and the parent nucleon, 
according to 
Refs.~\cite{Burkardt03,diehlhagler05} one defines three-dimensional 
densities representing the probability  to find a quark 
 with transverse spin $\tvec s_T$ in the
 nucleon with transverse spin $\tvec S_T$. It reads
\begin{eqnarray}
\rho(x,{\tvec b},{\tvec s}_T,{\tvec S}_T) 
= {}\dsp \oneh\left[ H(x,{b}^2)  + s^i_TS^i_T\left( H_T(x,{b}^2)  -\frac{1}{4M^2} \Delta_b \widetilde H_T(x,{b}^2) \right) \right.
\qquad
\qquad
\qquad
\qquad
\nonumber\\
%
+ \frac{b^j\varepsilon^{ji}}{M}\left(
S^i_T E'(x,{b}^2)  + s^i_T\left[ E'_T(x,{b}^2)  + 2 \widetilde H'_T(x,{b}^2) \right]\right)
\left.+ s^i_T(2b^ib^j - b^2\delta_{ij}) S^j_T\frac{1}{M^2} \widetilde H''_T(x,{b}^2) \right],
 \label{eq:transv}
 \end{eqnarray}
where the derivatives are defined
$
f' = \frac{\partial}{\partial b^2}\, f $, and
$
\Delta_b f
= 4\, \frac{\partial}{\partial b^2}
    \Big( b^2 \frac{\partial}{\partial b^2} \Big) f $.
In Eq.~(\ref{eq:transv}) enters the Fourier transform
of the GPDs in the impact-parameter space.
\newline
\noindent
In Eq.~(\ref{eq:transv}),
apart from the two orbitally symmetric monopole terms in the first line, 
there are two dipole structures proportional to
$b^j \eps^{ji} s_T^i$ and $b^j \eps^{ji} S_T^i,$
and a quadrupole term proportional to $s^i_T(2b^ib^j - b^2\delta_{ij}) S^j_T.$
The (derivatives of the) 
GPDs
$E(x,b^2)$, ${E}_T(x,b^2)+2\widetilde H_T(x,b^2)$ and  
$\widetilde{H}_{T}(x,b^2)$ 
thus determine how strongly the orbital symmetry in the transverse plane 
is distorted by the dipole and quadrupole terms. 

Lattice calculations accessing the lowest two $x$-moments of the quark
transverse-spin densities have recently been presented in Ref.~\cite{QCDSF06a}.
Here we show some results in the light-cone CQM 
for the first $x$-moment of the 
spin distributions in the cases of transversely polarized quarks in an unpolarized nucleon and unpolarized quarks in a transversely polarized nucleon, 
referring to~\cite{PBspin} 
for the discussions of more complex spin configurations.

\begin{figure}
\centerline{
  \includegraphics[width=12.5 cm]{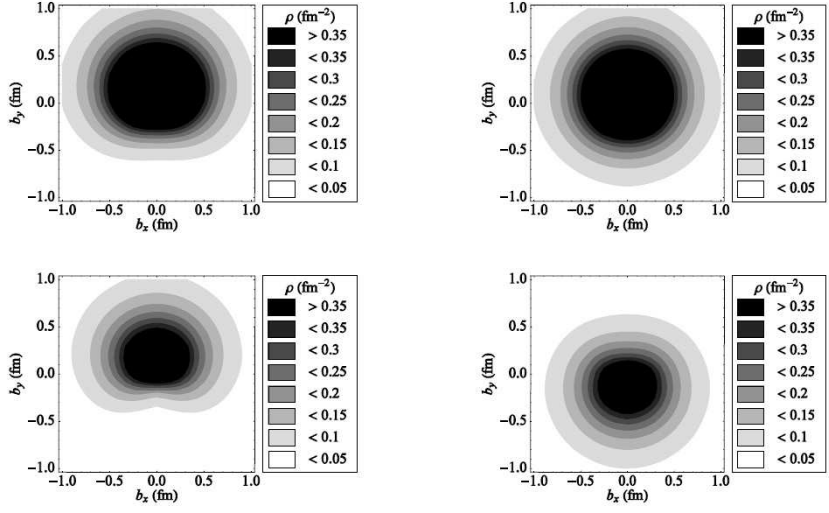}}
\caption{The spin-densities
for (transversely) $\hat x$-polarized quarks in an unpolarized proton
(left panels) and for unpolarized quarks in a  (transversely) $\hat x$-polarized proton.
The upper (lower) row corresponds to the results for up (down) quarks.}
\label{fig:fig1}
\end{figure}
In the case of transversely polarized quarks in an unpolarized proton 
the dipole 
contribution $E'_T(x,{b}^2)  + 2 \widetilde H'_T(x,{b}^2)$
introduces a large distortion perpendicular to both the quark 
spin and the momentum of the proton, as shown in the left column of
Fig.~\ref{fig:fig1}. 
Evidently, quarks in this situation also have a transverse component of 
orbital angular momentum. This effect has been 
related~\cite{Burkardt05b} to a non-vanishing Boer-Mulders
 function
$h_1^\perp$ which describes the correlation between 
intrinsic transverse momentum and transverse spin of quarks. 
Such a distortion reflects the large value of  the anomalous tensor magnetic
 moment $\kappa_T$ for both flavors. Here, $\kappa^u_T=3.98$ 
and $\kappa^d_T=2.60$, to be compared with the values $\kappa^u_T\approx 3.0$ 
and $\kappa^d_T\approx 1.9$ of Ref.~\cite{QCDSF06a}.
Since $\kappa_T\sim - h_1^\perp$, the present 
results confirm the conjecture that $h_1^\perp$ is large and negative both 
for up and down quarks~\cite{Burkardt05b}.

As also noticed in Refs.~\cite{Burkardt00a,QCDSF06a} the large anomalous 
magnetic moments $\kappa^{u,d}$ are responsible for the dipole distortion 
produced in the case of unpolarized quarks in transversely polarized
 nucleons (right column of Fig.~\ref{fig:fig1}). 
With the present model, $\kappa^u=1.86$ and $\kappa^d=-1.57$, 
to be compared with the values $\kappa^u=1.673$ and $\kappa^d=-2.033$ 
derived from data. This effect can serve as a dynamical explanation of
 a non-vanishing Sivers function
$f_{1T}^\perp$ 
which measures the correlation between the intrinsic quark transverse momentum 
and the transverse nucleon spin. The present results, with the opposite 
shift of up and down quark spin distributions, imply an opposite sign of 
$f_{1T}^\perp$ for up and down quarks~\cite{Burkardt02} as
 confirmed by the recent observation of the HERMES~\cite{Hermes05a} and 
COMPASS~\cite{COMPASS}
collaborations.
The results in Fig.~\ref{fig:fig1} are also in qualitative agreement with 
lattice calculations~\cite{QCDSF06a}.

\section{Energy-Momentum Tensor}

The nucleon EMT form factors were studied in lattice QCD~\cite{QCDSF04}, chiral perturbation theory~\cite{DMS}, and models such as the Skyrme model~\cite{CGOS07} and chiral quark
soliton model ($\chi$QSM)~\cite{GGOPSSU07,WN07}.
Here we summarize some interesting features investigated within the $\chi$QSM when
looking at the spatial distribution of the EMT form factors~\cite{Polyakov03,GGOPSSU07}.
The $\chi$QSM provides a field theoretic description of the nucleon in the limit of a large number of colors $N_c$, where the nucleon appears as chiral soliton of 
a static background field. Numerous nucleonic properties have been described 
in this model, 
giving predictions 
in agreement with phenomenology within an accuracy of (10-30)$\%$.
\newline
\begin{figure}[b]
\begin{center}
\begin{tabular}{cc}
\includegraphics[width=14cm]{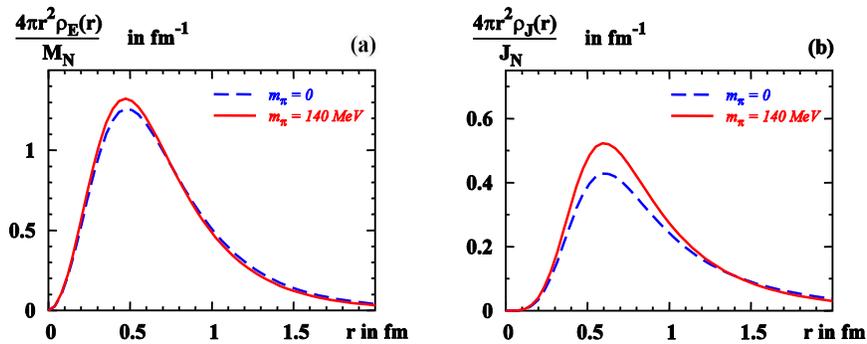}
\end{tabular}
\end{center}
\caption{\small (a) The normalized energy density $4\pi r^2\rho_E(r)/M_N$ from the $\chi$QSM as a function of $r$ in the chiral limit of $m_\pi=0$ (dashed curve) and for $m_\pi=140\,{\rm MeV}$ (solid curve). (b) The same for the normalized angular momentum density $4\pi r^2\rho_J(r)/J_N$ (taken from~\cite{GGOPSSU07}).}
\label{fig:enspin-density}
\end{figure}
The spatial distribution of the EMT form factors
can be obtained from
the Fourier transform  with respect to ${\boldmath \Delta}$
of the matrix element of the static EMT 
calculated in the Breit frame~\cite{Polyakov03}.
The normalized energy density $4\pi r^2\rho_E(r)/M_N$ 
is shown in Fig.~\ref{fig:enspin-density}a as a function of $r$ in the 
chiral limit of a vanishing pion mass and for the physical situation with a 
pion mass of $140\,{\rm MeV}$. 
In the center of the nucleon the energy density is $\rho_E(0)=1.70$ GeV/fm$^3$ 
or $3.0\times 10^{15}$ g cm$^{-3}$, 
corresponding roughly to 13 times the equilibrium density of nuclear matter. 
As the pion mass decreases, the energy density is spread more widely. 
According to the role of the pion field in the $\chi$QSM, where one can 
associate the contribution of the discrete level to the quark core and the 
contribution of the negative continuum states to a pion cloud, 
this means that the range of the pion cloud increases and the nucleon becomes 
larger. 
Actually, the mean square radius $\langle r_E^2\rangle$ increases from 
0.67 fm$^2$ in the case of the physical pion to 0.79 fm$^2$ 
in the chiral limit. 
With increasing pion mass up to 1.2 GeV this trend is confirmed with the 
nucleon becoming smaller and smaller~\cite{GGOPSSU07}.

\begin{figure}[t]
\begin{center}
\begin{tabular}{lll}
    \includegraphics[height=4.5cm]{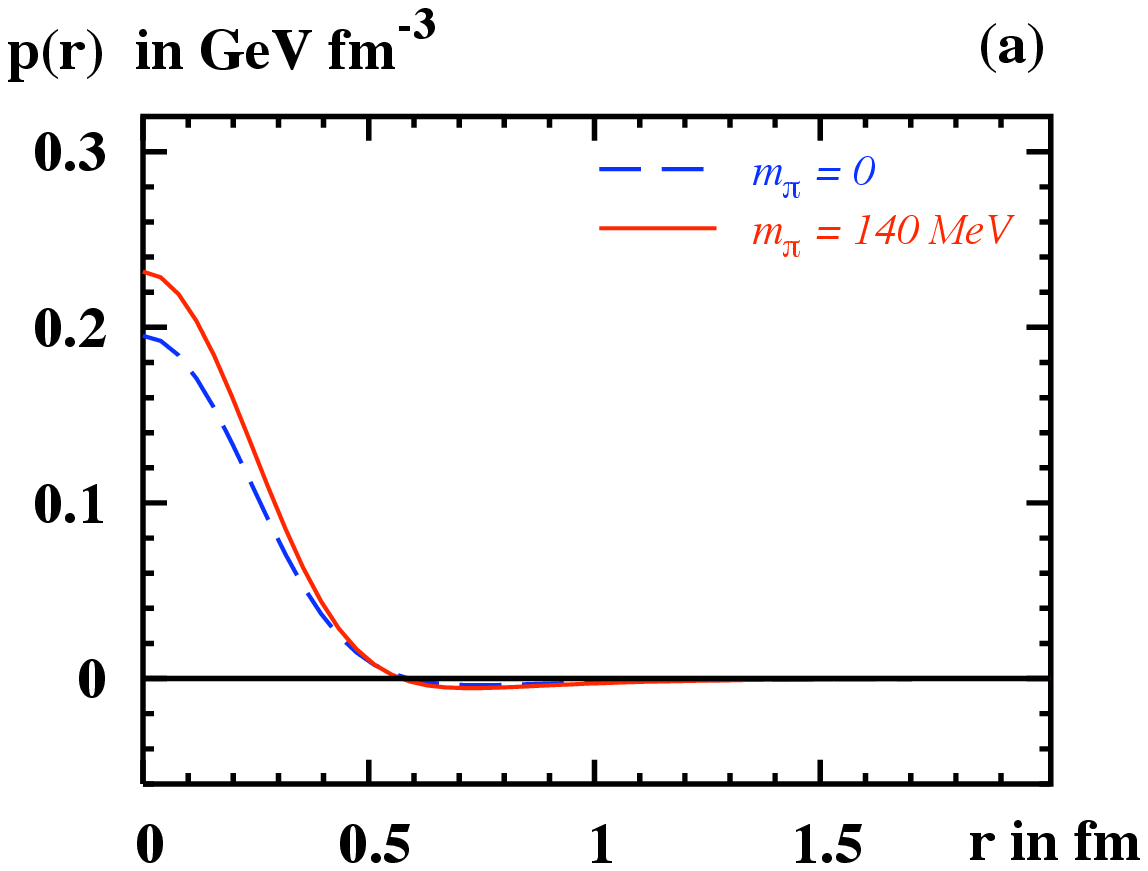}&
    \includegraphics[height=4.5cm]{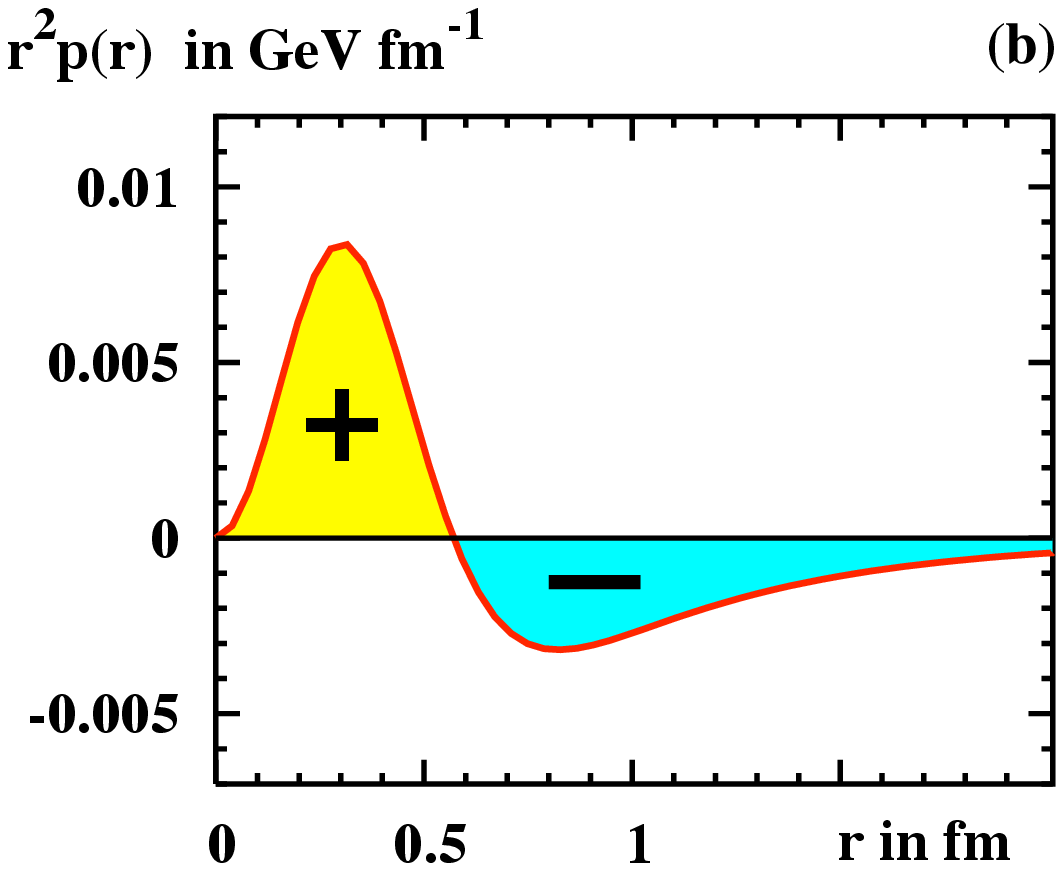} 
\end{tabular}
\end{center}
    \caption{\small (a) The pressure $p(r)$ from the $\chi$QSM as function of $r$ for $m_\pi=0$ (dashed curve) and $140\,{\rm MeV}$ (solid curve). (b) $r^2 p(r)$ as function of $r$ at the physical value of $m_\pi$ (taken from~\cite{GGOPSSU07}).}
\label{fig:pressure}
\end{figure}

The normalized 
radial distribution of angular momentum $4\pi r^2\rho_J(r)/J_N$, 
is shown in Fig.~\ref{fig:enspin-density}b as a function of $r$ for $m_\pi=0$ 
and $140\,{\rm MeV}$. 
For any $m_\pi$ at small $r$ one finds $\rho_J(r) \propto r^2$. 
The mean square radius $\langle r_J^2\rangle$ decreases with increasing 
$m_\pi$~\cite{GGOPSSU07} in agreement with the idea of a shrinking pion cloud. 
For a physical pion one finds $\langle r_J^2\rangle= 1.32$ fm$^2$. 
At large $r$ in the chiral limit $\rho_J(r) \propto 1/r^4$ such 
that $\langle r_J^2\rangle$ diverges.
\newline
Fig.~\ref{fig:pressure}a shows the pressure $p(r)$ as function of $r$. 
In the physical situation $p(r)$ takes its global maximum at $r=0$ with 
$p(0) = 0.23\,{\rm GeV}/{\rm fm}^3 = 3.7\cdot10^{34}\,{\rm Pa}$. 
This is ${\cal O}(10\!-\!100)$ higher than the pressure inside a neutron star. 
Then $p(r)$ decreases monotonically (becoming zero at $r_0 = 0.57\,{\rm fm}$) 
till reaching its global minimum at 
$r_{p,\,\rm min}= 0.72\,{\rm fm}$, after which it increases monotonically 
remaining, however, always negative. 
The positive sign of the pressure for $r<r_0$ corresponds to the repulsion 
among quarks imposed by Pauli principle, while the negative sign in the region
 $r > r_0$ means attraction in agreement with the idea of a pion cloud 
responsible for binding the quarks to form the nucleon. 
The subtle balance between repulsion and attraction, ultimately producing 
a stable soliton, can be better appreciated from Fig.~\ref{fig:pressure}b 
showing $r^2p(r),$ where
the shaded regions have the same surface areas but opposite sign and cancel each other within numerical accuracy.


\end{document}